\documentclass[amsmath, amssymb, 12pt]{article}
\usepackage{epsfig}
\def\be{\begin{equation}}
\def\ee{\end{equation}}

\usepackage[a4paper,vmargin={33mm,37mm},hmargin={25mm,20mm}]{geometry}

\long\def\@makefntext#1{\parindent 0cm\noindent \hbox to
1em{\hss$^{\@thefnmark}$}#1}

\begin{document}

\begin{titlepage}
\vspace{.5in}
\begin{flushright}
\end{flushright}
\vspace{.5in}
\begin{center}
{\Large\bf   Gravitational Lensing by Phantom Black holes}\\
\vspace{.4in}{Galin N. Gyulchev $^{1}$\footnote{\it email: gyulchev@phys.uni-sofia.bg},
               Ivan Zh. Stefanov$^{2}$\footnote{\it email: izhivkov@tu-sofia.bg}\\

       { \footnotesize  ${}^{1}$ \it Department of Physics, Biophysics and Roentgenology, Faculty of Medicine, Snt. Kliment Ohridski University of Sofia, 1, Kozyak Str., 1407 Sofia, Bulgaria  }\\

       {\footnotesize \it ${}^{2}$ Department of Applied Physics, Technical University of Sofia, 8, Snt. Kliment Ohridski Blvd., 1000 Sofia, Bulgaria}}

\end{center}

\vspace{.5in}

\begin{center}
{\large\bf Abstract}
\end{center}
In some models dark energy is described by phantom scalar fields (scalar fields with "wrong" sign of the kinetic term in the lagrangian). In the current paper we study the effect of phantom scalar field and/or phantom electromagnetic field on gravitational lensing by black holes in the strong deflection regime. The black-hole solutions that we have studied have been obtained in the frame of the Einstein--(anti--)Maxwell--(anti--)dilaton theory. The numerical analysis shows considerable effect of the phantom scalar and electromagnetic fields on the angular position, brightness and separation of the relativistic images. \\ \, \\



\end{titlepage}
\addtocounter{footnote}{-1}

\section{Introduction}

Modern observational programs including  type Ia SNe, cosmic microwave background aniso\-tropy and
mass power spectrum suggest that the universe is dominated by mysterious matter termed dark energy (DE) which has negative pressure and violates the energy conditions \cite{Tonryet, Hannestad, Jarosik}. Considerable efforts are made to study the nature of DE.
Different effective models of dark energy have been proposed in literature (See \cite{reviewDE} and \cite{Tsujikawa} for recent exhaustive reviews). In some of them the possibility of describing DE by phantom fields is considered.

The natural questions arises whether local manifestations of DE at astrophysical scale can be observed. Exact solutions describing neutron stars containing DE have been obtained in \cite{Stoytcho}. There have been also some recent efforts in that direction. In \cite{DanielaStoytcho} the effect of DE on the structure and on the spectrum of qusinormal frequencies of neutron stars has been studied.  Mixed stars containing both dark energy and ordinary matter have been presented in a number of papers (See \cite{Stoytcho} and references therein). Solutions describing black holes coupled to phantom fields have also been found. To our knowledge the first solutions of phantom black holes have been obtained by Gibbons and Rasheed \cite{GibbonsRasheed}. These solutions were later elaborated by Cl\'{e}ment et al. \cite{Clement, Clement_sigma} and Gao Zhang \cite{GaoZhang} for higher dimensions. Regular black holes coupled to phantom scalar field have been reported by Bronnikov \cite{Bronnikov}. Recent interest in phantom black holes have been connected with the study of their thermodynamics and the possibility of phase transitions \cite{Rodrigues1}. Similar study has been presented in \cite{Rodrigues2} for black holes with phantom electromagnetic field or the so-called anti-Reissner Nordstr\"{o}m black hole. In this solution the charged term in the metric has an opposite sign with respect to the corresponding term of the standard Reissner Nordstr\"{o}m black hole. Other works in the field of theories with phantom dilaton and phantom Maxwell field  have considered gravitational collapse of a charged scalar field \cite{phantom_collapse} and  also  light paths in black-hole space-tims \cite{phantom_light_path}.

As we have already mentioned gravitational waves and the frequencies of quasinormal ringing in particular can provide rich information for the structure of compact astrophysical objects and thus can serve as a powerful tool for studying  the local manifestation of DE. Another possibility could be provided by gravitational lensing especially in the strong deflection regime. There has been considerable effort for the theoretical study of gravitational lensing in the strong deflection regime (For more details on the matter we refer the reader to \cite{KeetonPetters} and references therein). In his papers \cite{Bozza1, Bozza2} Bozza proposed a method for the calculation of the deflection angle in the regime of strong deflection in the particular case when both the observer and the gravitational source lie in the equatorial plane\footnote{One should mention, however, that the precision of Bozza's method has been questioned by Virbhadra in his paper \cite{VirbhadraRelativistic}.} . His method has gained popularity due to is simplicity and has been applied to study the gravitational lensing caused by different exotic, compact objects. The particular cases in which both the scalar field and the electromagnetic field have cannonical form, i.e. the EMD black hole has been already reported by Bhadra \cite{Bhadra2003}. The lensing by EMD black holes with de-Sitter and anti-de-Sitter asymptotics have been studied by \cite{dilatonDeSitter} and \cite{Sengupta}, respectively. In the last two cases the scalar field has a non zero potential. Lensing in the strong field regime by black holes coupled to electromagnetic field has been considered also in \cite{Whisker, EBI, GyulchevYazadjiev1, Sendra, DingJing, chargedKK}.

Black holes with opposite sign of the charge term in the  metric (as in the case of anti-Reissner Nordstr\"{o}m black hole) have been applied to model the object in the center of our galaxy --  Sgr A* and their lensing has been studied in \cite{SgrA} and \cite{Horvath}. In these black holes, however, the charge is tidal and does not have electromagnetic origin. Lensing by black holes with tidal charge gas been also considered in \cite{Randall}.

One of the aims of the current paper is to study the effect of phantom scalar field (phantom dilaton) on gravitational lensing. In the presence of exotic matter such as phantom fields wormholes may exist. Lensing by different wormholes, for example the  Ellis's and the Janis-Newman-Winicour's (JNW) wormholes, has attracted significent research interest \cite{wormholeClement}--\cite{Harada}. JNW naked singularities (naked singularities coupled to canonical massless scalar field) acting as gravitational lens have been considered by Virbhadra et al. \cite{role_of_the, VE3, VK}.
The lensing of the JNW solution in the context of scalar-tensor theories has been studied by Bhadra \cite{Bhadra}. Generalization with inclusion of rotation has been made in \cite{GyulchevYazadjievRNS}.

Our goal is apply the apparatus of gravitational lensing by black holes in the strong deflection limit to study the possible local manifestation of dark energy. For this purpose we model DE with phantom dilaton and phantom electromagnetic field. We compare the characteristics of relativistic images of four black holes: the standard Einstein-Maxwell black hole (EMD); the Einstein-anti-Maxwell-dilaton black hole which has a phantom electromagnetic field (E$\overline{\rm M}$D)\footnote{We will adopt the abbreviations introduced in \cite{GibbonsRasheed}. };  the Einstein-Maxwell-anti-dilaton black hole which has a phantom dilaton (EM$\overline{\rm D}$); and  the Einstein-anti-Maxwell-anti-dilaton black hole in which both the dilaton and the electromagnetic field are phantom (E$\overline{\rm M}\overline{\rm D}$).

\section{Phantom black holes}

When phantom dilaton and/or phantom electromagnetic field is considered the action of Einstein-Maxwell-dilaton theory is generalized to the following form
\be
S=\int dx^{4}\sqrt{-g}\left[  R-2\,\eta_{1}
g^{\mu\nu}\nabla_{\mu}\varphi\nabla_{\nu }\varphi+\eta_{2} \,e^{
-2\alpha\varphi}F^{\mu\nu}F_{\mu\nu}\right].  \label{action1}\;
\ee
$R$ denotes the Ricci scalar curvature, $\varphi$ is the dilaton, $F$ is the Maxwell tensor and the constant $\alpha$ determines the coupling between the dilaton and the electromagnetic field. For the usual dilaton the dilaton-gravity coupling constant $\eta_1$ takes the value $\eta_{1}=1$ while for phantom dilaton $\eta_{1}=-1$. Similarly, the
Maxwell-gravity coupling constant $\eta_{2}$ takes the values $\eta_{2}=1$ and $\eta_{2}=-1$ in the Maxwell and anti-Maxwell case, respectively.

\subsection{Einstein Maxwell Dilaton black holes}

The line element of the EMD black hole\footnote{This is the so-called Gibbons-Maeda-Garfinkle-Horowitz-Str\"{o}minger black-hole solution \cite{GM, GHS}.} is
$$
ds^2 = -\left(1-{r_+\over r}\right) \left(1-{r_-\over
r}\right)^{\gamma}dt^2 + \left(1-{r_+\over r}\right)^{-1}
\left(1-{r_-\over r}\right)^{-\gamma}dr^2
$$
\be
+r^2 \left(1-{r_-\over r}\right)^{1-\gamma}(d\theta^2 +
\sin^2\theta d \phi^2),\label{metric_EMD}
\ee
where the parameter $\gamma=(1-\alpha^2)/(1+\alpha^2)$ has been introduced for convenience. It varies in the interval $[-1,1]$ for $\alpha\in(-\infty,\infty)$, so stronger coupling corresponds to lower values of $\gamma$.
The solutions for the dilaton and the Maxwell field are
\be
e^{2\alpha\varphi} = \left(1-{r_-\over r}\right)^{1-\gamma}, \quad\quad F = {Q\over r^2} dt\wedge dr
\ee
For the magnetically charged solution the metric is the same but the sign of the
scalar field $\varphi$ must be reversed and the Maxwell field becomes
$F=P\sin\theta d\theta\wedge d\phi$.
The parameters  $r_+$ and $r_-$ are interpreted as an event horizon and an inner Cauchy horizon, respectively. The ADM mass $M$  and the charge $Q$ can be expressed by $r_+$ and $r_-$
\be
2M=r_++\gamma r_-, \quad\quad 2Q^2=(1+\gamma)r_+r_-. \label{charge_mass}
\ee
Relations (\ref{charge_mass}) can be inverted to express the horizons in terms of the ADM mass $M$  and the charge $Q$
\be
r_+=M\left[1+\sqrt{1-{2\gamma \over 1+\gamma}\left({Q\over M}\right)^2}\right],\quad r_-={M\over\gamma}\left[1-\sqrt{1-{2\gamma \over 1+\gamma}\left({Q\over M}\right)^2}\right]
\ee
The equation for $r_{+}$ (or $r_{-}$) obtained from (\ref{charge_mass}) is biquadratic. The solutions are grouped in two couples. The couple which contains the largest of all four roots is chosen. The same choice is made in the other three classes of solutions considered in this paper.
The two horizons merge at
\be
\left({Q\over M}\right)^2=\left({Q\over M}\right)^2_{{\rm crit}}={2\over 1+\gamma}
\ee
and for lower values of $(Q/M)^2$ the solution describes a naked singularity. In the limit $\gamma\rightarrow1$ the solution restores the Reissner-Nordstr\"{o}m black hole.
The charge is switched off when one of the two parameters $r_{+}$ and $r_{-}$ is equal to zero. In the latter case, the Schwarzschild black hole is recovered with $r_{+}=2M$ corresponding to the event horizon. In the former case, the EMD solution reduces to the Janis-Newman-Winicour solution also known as the Fisher solution -- a fact that was noticed for the first time by Virbhadra \cite{VirbhadraWyman}. In this case, at $r_{-}=2M/\gamma$ a singularity is reached and $\gamma\in[0, 1]$. In the current work we will restrict our considerations to gravitational lensing of black holes. That is why we have chosen the Schwarzschild black hole as a reference. The gravitational lensing by the central object of the JNW spacetime has been studied in \cite{role_of_the, VE3}.

\subsection{Einstein anti-Maxwell Dilaton black holes}

In the case of E$\overline{\rm M}$D black hole the line element is again (\ref{metric_EMD}).
The solutions for the dilaton and the anti-Maxwell field are
\be
e^{2\alpha\varphi} = \left(1-{r_-\over r}\right)^{1-\gamma}, \quad\quad F = -{Q\over r^2} dt\wedge dr
\ee
The ADM mass $M$  and the anticharge $Q$ are
\be
2M=r_++\gamma r_- , \quad\quad  2Q^2=-(1+\gamma)r_+r_-.
\ee
The ``horizons'' expressed in terms of the ADM mass $M$  and the anticharge $Q$ are
\be
r_+=M\left[1+\sqrt{1+{2\gamma \over 1+\gamma}\left({Q\over M}\right)^2}\right],\quad r_-={M\over\gamma}\left[1-\sqrt{1+{2\gamma \over 1+\gamma}\left({Q\over M}\right)^2}\right]
\ee
The parameter  $r_+$ is positive and is interpreted as an event horizon while $r_-$ is a negative and can be considered as a singularity which is never reached since the singularity at $r=0$ is reached before that. Hence, these black holes have the same causal structure as the Schwarzschild black hole. Again, there is restriction for the parameter $(Q/M)$
\be
\left({Q\over M}\right)^2\leq\left({Q\over M}\right)^2_{{\rm crit}}=-{1+\gamma\over 2\gamma}.
\ee
 The limit $\gamma\rightarrow1$ corresponds to the anti-Reissner-Nordstr\"{o}m black hole (a Reissner-Nordstr\"{o}m black hole black hole with imaginary charge). $\left(Q/M\right)$ is unbound for positive $\gamma$. Again, the particular solutions with zero electric charge are the Janis-Newman-Winicour solution and the Schwarzschild solution.

\subsection{Einstein Maxwell anti-Dilaton black holes}

The line element of the EM$\overline{\rm D}$ black hole is
$$
ds^2 = -\left(1-{r_+\over r}\right) \left(1-{r_-\over
r}\right)^{1/\gamma}dt^2 + \left(1-{r_+\over r}\right)^{-1}
\left(1-{r_-\over r}\right)^{-1/\gamma}dr^2
$$
\be
+r^2 \left(1-{r_-\over r}\right)^{1-{1/\gamma}}(d\theta^2 +
\sin^2\theta d \phi^2),\label{metric_EM_D}
\ee
The solutions for the dilaton and the Maxwell field are
\be
e^{2\alpha\varphi} = \left(1-{r_-\over r}\right)^{1-1/\gamma}, \quad\quad F = {Q\over r^2} dt\wedge dr
\ee
When $\gamma>0$, $0\leq r_-\leq r_+$, so the causal structure is the same as for the EMD case. For $\gamma<0$, however,  $ r_-\leq 0\leq r_+$ and the black hole has the same causal structure as in the E$\overline{\rm M}$D case.  The ADM mass $M$  and the charge $Q$ are expressed by $r_+$ and $r_-$ in the following way
\be
2M=r_++{1\over\gamma} r_-, \quad\quad 2Q^2={(1+\gamma)\over\gamma}r_+r_-.\label{charge_mass_EM_D}
\ee
Relations (\ref{charge_mass_EM_D}) can be inverted to express the ``horizons'' in terms of the ADM mass $M$  and the charge $Q$
\be
r_+=M\left[1+\sqrt{1-{2 \over 1+\gamma}\left({Q\over M}\right)^2}\right],\quad r_-={\gamma M}\left[1-\sqrt{1-{2 \over 1+\gamma}\left({Q\over M}\right)^2}\right].
\ee
For $r_+$ and $r_-$ to be real the following relation must hold
\be
\left({Q\over M}\right)^2\leq\left({Q\over M}\right)^2_{{\rm crit}}={1+\gamma\over 2}.
\ee
Here in the limit $\gamma\rightarrow1$ the Reissner-Nordstr\"{o}m black hole is restored.
For $r_-=0$ the Schwarzschild black hole is restored. If we put $r_+=0$ and substitute $\gamma=1/\kappa$ the metric obtains the form
$$
ds^2 = - \left(1-{r_-\over r}\right)^{\kappa}dt^2 + \left(1-{r_-\over r}\right)^{-\kappa}dr^2+r^2 \left(1-{r_-\over r}\right)^{1-{\kappa}}(d\theta^2 +
\sin^2\theta d \phi^2).\label{antiJNWmetric}
$$
This is the anti-Fisher or anti-JNW solution since $\kappa\in[-1, \infty)$. Lensing in this spacetime has been studied in \cite{Sen}.
\subsection{Einstein anti-Maxwell anti-Dilaton black holes}

In the case of E$\overline{\rm M}$$\overline{\rm D}$ black hole the line element is given again by (\ref{metric_EM_D}).
The solutions for the dilaton and the anti-Maxwell field are
\be
e^{2\alpha\varphi} = \left(1-{r_-\over r}\right)^{1-1/\gamma}, \quad\quad F = -{Q\over r^2} dt\wedge dr
\ee
When $\gamma>0$, $r_-\leq0\leq  r_+$ and the causal structure is Schwarzschild-like. For $\gamma<0$, however,  $  0\leq r_-\leq r_+$ and the black hole has two horizons, an event horizon and an inner Cauchy horizon.
The ADM mass $M$  and the anticharge $Q$ are
\be
2M=r_++{1\over\gamma} r_-, \quad\quad 2Q^2=-{(1+\gamma)\over\gamma}r_+r_-.\label{charge_mass_E_M_D}
\ee
Relations (\ref{charge_mass_E_M_D}) can be inverted to express the``horizons'' in terms of the ADM mass $M$  and the charge $Q$
\be
r_+=M\left[1+\sqrt{1+{2 \over 1+\gamma}\left({Q\over M}\right)^2}\right],\quad r_-={\gamma M}\left[1-\sqrt{1+{2 \over 1+\gamma}\left({Q\over M}\right)^2}\right].
\ee
Unlike all three cases discussed above in the current case there are no restrictions for $(Q/M)^2$.
The limit $\gamma\rightarrow1$ corresponds, again, to the anti-Reissner-Nordstr\"{o}m black hole.
The particular solutions with zero electric charge are the anti-JNW solution and the Schwarzschild solution.
\section{Gravitational lensing in the strong field limit}

Following Bozza's notation we can express the metric of the general static spherically symmetric spacetime in the form

\begin{equation}
ds^{2}=A(x)dt^{2}-B(x)dx^{2}-x^{2}(d\theta^{2}+\sin^{2}\theta
d\varphi^{2})
\end{equation}
where we have introduced the new variable $x=r/M$.
\noindent The deflection angle can be expressed as \cite{role_of_the}

\begin{equation}
\alpha(x_{0})=I(x_{0})-\pi
\end{equation}

\noindent where

\begin{equation}
I(x_{0})=2\int_{x_{0}}^{\infty}\frac{\sqrt{B(x)}}{\sqrt{C(x)}
\sqrt{\frac{C(x)A(x_{0})}{C(x_{0})A(x)}}-1}dx\label{integral}
\end{equation}

\noindent and here $x_{0}$ represents the minimum distance from
the photon trajectory to the gravitational source.
The deflection angle  diverges when the
denominator of the above expression turns to zero i.e. at the points where the following relation
$\frac{C\,'(x)}{C(x)}=\frac{A'(x)}{A(x)}$  holds. We use prime $(..)'$ to denote the derivative with respect to $x$. The largest root of this equation gives the radius of the photon sphere. For more details on photon surfaces we refer the reader to \cite{VE2, CVE3, VE3}

Here and bellow the following convention has been chosen
$
F_{m}=F|_{r_{0}=r_{m}}
$
where $F$ is an arbitrary quantity.

Acoording to Bozza's method \cite{Bozza1, Bozza2} the integral (\ref{integral}) is split in two parts -- regular $I_{R}(x_{0})$ and divergent $I_{D}(x_{0})$
\begin{equation}
I(x_{0})=I_{D}(x_{0})+I_{R}(x_{0}).
\end{equation}
In explicit form
\begin{equation}
I_{D}(x_{ps})=\int_{0}^{1}\frac{u_{ps}}{\sqrt{\beta_{ps}}}\sqrt{\frac{B_{ps}}{C_{ps}}}\frac{x_{ps}}{\eta}d\eta,
\end{equation}
\begin{equation}\label{IR}
I_{R}(x_{ps})=\int_{0}^{1}\left [ u_{ps}\sqrt{\frac{B(\eta)}{C(\eta)}}[R(\eta,u_{ps})]^{-1/2}\frac{x_{ps}}{(1-\eta)^2}-\frac{u_{ps}}{\sqrt{\beta_{ps}}}\sqrt{\frac{B_{ps}}{C_{ps}}}\frac{x_{ps}}{\eta}\right ] d\eta.
\end{equation}
In this formulas the following quantities have been introduced. The new variable
\begin{equation}
\eta=1-\frac{x_{ps}}{x_{0}},
\end{equation}
facilitates the numerical integration since it maps the open interval $[x_{ps}, \infty)$ to the closed interval $[0,1]$.
The function
\begin{equation}
R(\eta,u_{ps})=\frac{C(\eta)}{A(\eta)}-u_{ps}^2
\end{equation}
is responsible for the divergence of the integrand. As the photon sphere is approached, i.e. when $\eta\rightarrow0$ the leading order term  of the integrand is $(\sqrt{\beta_{ps}}\eta)^{-1}$. The coefficient in the expansion is
\begin{equation}
    \beta_{ps}=\frac{1}{2}x_{ps}^2\frac{C_{ps}^{\prime\prime}A_{ps}-C_{ps}A_{ps}^{\prime\prime}}{A_{ps}^2}.
\end{equation}
The expansion shows that divergence of the deflection angle is logarithmic \cite{Bozza1, Bozza2}
\begin{equation}
\alpha(\theta)=-a\ln\left(\frac{\theta
D_{OL}}{u_{ps}}-1\right)+b+O(u-u_{ps}).
\end{equation}
\noindent where $D_{OL}$ denotes the distance between observer and
gravitational lens. The impact parameter is
\begin{equation}
u_{ps}=\sqrt{\frac{C_{ps}}{A_{ps}}}.
\end{equation}
\noindent The strong field limit coefficients $a$ and
$b$ are expressed as,
\begin{equation}
a=x_{ps}\sqrt{\frac{B_{ps}}{A_{ps}\beta_{ps}}},
\end{equation}
\begin{equation}
b=-\pi+I_{R}(x_{ps})+a\ln\left(\frac{2\beta_{ps}}{u_{ps}^2}\right).
\end{equation}

Since the spacetimes under consideration are asymptotically flat we can take advantage of the strong deflection limit lens equation \cite{Bozza3}
\begin{equation}
    \eta=\frac{D_{OL}+D_{LS}}{D_{LS}}\theta-\alpha(\theta) \hspace{0.3cm} \rm{mod} \hspace{0.3cm} 2\pi,  \label{LensEquation}
\end{equation}
where $D_{LS}$ is the lens-source distance, $D_{OL}$ is the observer-lens distance and $\eta$ is the source angular position, as seen from the lens. We will be interested also in the following observables. Under the assumption $u_{ps}\ll D_{OL}$, one can show that up to terms of second order in $u_{ps}/D_{OL}$ the angular separation between the lens and the \emph{n}-th relativistic image is
\begin{equation}¶
\theta_n^{pro}=\theta_n^{0}\left(1-{u_{ps}e_n^{pro}(D_{OL}+D_{LS})\over a D_{OL}D_{LS}}\right),
\end{equation}¶
where
\begin{equation}¶
\theta_n^{0}={u_{ps}\over D_{OL}}\left(1+e_n^{pro}\right),\quad\quad e_n^{pro}=e^{b-|\eta|+2\pi n\over  a}.
\end{equation}¶
We are considering only prograde photons and this is what $pro$ stands for.
It is usually considered that only the first relativistic image can be observed separately and all other relativistic images would be packed together at angular position $\theta_{\infty}$. The angular separation between the first relativistic image and the rest of the relativistic images is \cite{Bozza1}
\begin{equation}¶
s_1^{pro}=\theta_{1}-\theta_{\infty}=\theta_{\infty}e^{b-2\pi\over  a}.
\end{equation}¶
The third observable that is usually considered is the ratio between the magnitude of the first image $\mu_1$ and the total magnitude of all other relativistic images $\sum_{n=2}^{\infty}\mu_n$
\begin{equation}¶
r={\mu_1\over\sum_{n=2}^{\infty}\mu_n}=e^{2\pi\over  a},
\end{equation}¶
which in terms of stellar magnitudes is
\begin{equation}¶
r_m=2.5\lg(r).
\end{equation}¶

All observable quantities mentioned above are plotted in the paper for different values of the charge $Q/M$ and the metric parameter $\gamma$ and under the following assumptions. We consider the massive dark object Sgr $\rm A^{\ast}$ in the center of our Galaxy as a lens. The observer is positioned at distance $D_{OL}=8.33$ kpc from the lens. For the lens-source distance, following \cite{VirbhadraRelativistic}, we have taken  $D_{LS}=0.005D_{OL}$, $D_{LS}=0.05D_{OL}$ and $D_{LS}=0.5D_{OL}$. According to \cite{BHMass} the lens mass is $M=4.31\times10^{6}M_{\odot}$, so $M/D_{OL}\approx2.47\times10^{-11}$. As in the Schwarzschild case\footnote{See Fig. 3 in \cite{VirbhadraRelativistic}} our results are practically insensitive to the angular source position $\eta$, and the source distance $D_{LS}$ . For simplicity we will present the results for $\eta=0$. In this specific case the relativistic images are observed as Einstein rings \cite{VE3}.

We should also mention that significant information about the properties of the object acting as a gravitational lens can be obtained from the time delay, however such study is not in the scope of the present work. Expression for the time delay in general static spherically symmetric spacetime can be found in \cite{VK}.
\subsection{Photon sphere}

\begin{figure}[b]%
\begin{center}
\includegraphics[width=0.4\textwidth]{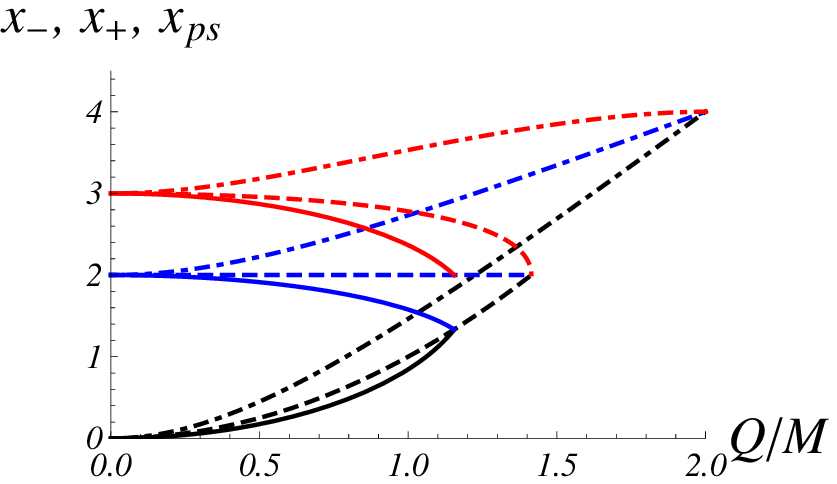}
\includegraphics[width=0.4\textwidth]{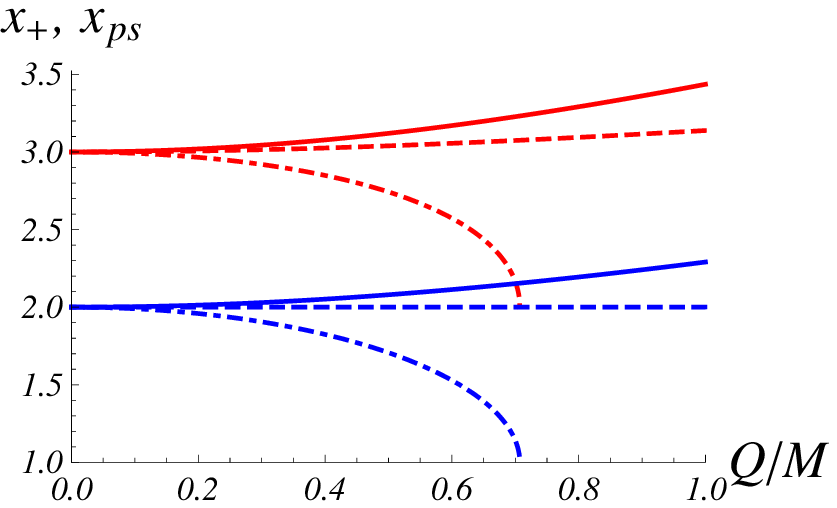}
\caption{\small
The photon sphere $x_{ps}$ (red), the event horizon  $x_{+}$  (blue) and the inner horizon $x_{-}$ (black) of the EMD and E${\rm \overline{M}}$D black
holes for three values of $\gamma$: $\gamma=-0.5$ (dash-dot), $\gamma=0$ (dash) and $\gamma=0.5$ (solid).} \label{xps_EMD_E_MD}%
\end{center}
\end{figure}%

For both solutions with canonical scalar field, ${\rm EMD}$ and ${\rm E\overline {M }D}$, the expression for the photon sphere is
\begin{equation}¶
    x_{ps}=\frac{3}{4}x_{+}+\frac{1}{4}\left( 2\gamma+1 \right) x_{-}+
\frac{1}{4}\sqrt{9{x_{+}}^{2}+ \left( 2\gamma+1 \right)^{2}{x_{-}}^{2}-2\left( 2\gamma+5 \right) x_{+}x_{-}},
\end{equation}¶
where $x_{+}=r_{+}/M$ and $x_{-}=r_{-}/M$, $r_{+}$ and $r_{-}$ are the parameters of the corresponding black-hole solution. The photon sphere $x_{ps}$, the event horizon  $x_{+}$ and the inner horizon $x_{-}$ of the EMD and E${\rm \overline{M}}$D black holes are displayed on Fig. {\ref{xps_EMD_E_MD}}. In the EMD case for $\gamma<0$ the photon sphere and the event horizon merge when $\left(Q/ M\right)=\left(Q/ M\right)_{{\rm crit}}$. This situation has been recently discussed in \cite{EMD_extremal}. In the ${\rm E\overline{M}D}$ case we can see that  $\left(Q/ M\right)$ is restricted from above only when $\gamma<0$. The photon sphere and the event horizon do not merge for any value of $\left(Q/ M\right)$ in this case. The inner "horizon" $x_{-}$ is behind the central singularity and is not present on the figure.

\begin{figure}[t]%
\begin{center}
\includegraphics[width=0.4\textwidth]{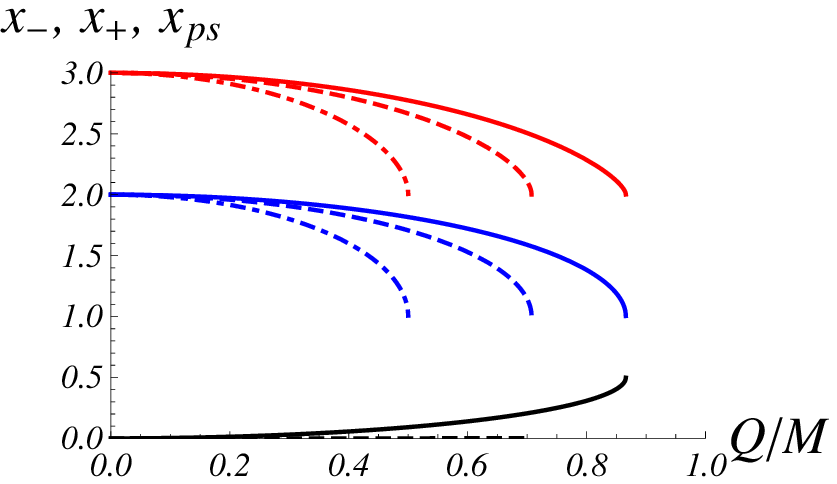}
\includegraphics[width=0.4\textwidth]{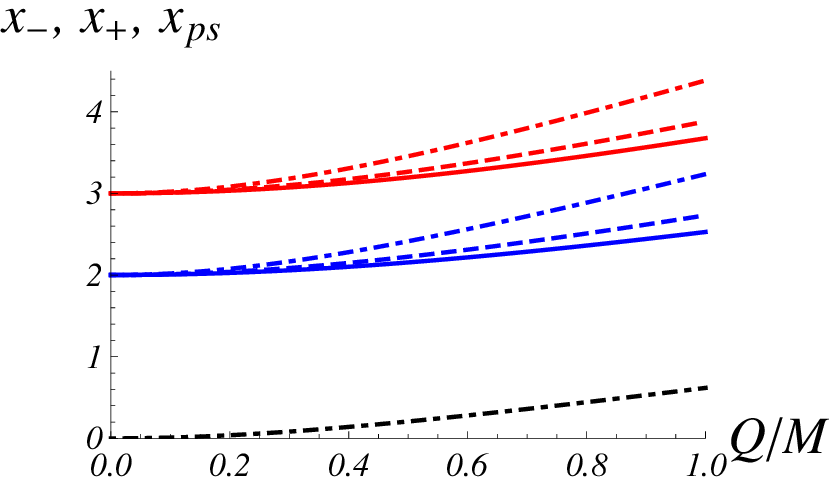}
\caption{\small The photon sphere $x_{ps}$ (red), the event horizon  $x_{+}$ (blue) and the inner horizon $x_{-}$ (black) of the EM${\rm \overline{D}}$ and  E${\rm \overline{M}}$${\rm \overline{D}}$ black holes for three values of $\gamma$:
$\gamma=-0.5$ (dash-dot), $\gamma=0$ (dash) and $\gamma=0.5$ (solid).} \label{xps_EM_D_E_M_D}%
\end{center}
\end{figure}

For the solutions with phantom scalar field, ${\rm EM\overline{D}}$ and ${\rm E\overline{\rm M}\overline{D}}$, the photon sphere takes the form
\begin{equation}¶
    x_{ps}=\frac{3}{4}x_{+}+\frac{1}{4}\left( \frac{2}{\gamma}+1 \right)x_{-}+
\frac{1}{4}\sqrt{9{x_{+}}^{2}+ \left( \frac{2}{\gamma}+1 \right)^{2}{x_{-}}^{2}-2\left( \frac{2}{\gamma}+5 \right) x_{+}x_{-}},
\end{equation}¶
The photon sphere $x_{ps}$, the event horizon  $x_{+}$ and the inner horizon $x_{-}$ of the EM${\rm \overline{D}}$ and E${\rm \overline{M}}$${\rm \overline{D}}$ black holes are displayed on Fig. {\ref{xps_EM_D_E_M_D}}.
In the EM${\rm \overline{D}}$ case $\left(Q/ M\right)$ is restricted from above. When $\gamma<0$ the inner "horizon" $x_{-}$ is behind the central singularity and is not present on the figure. There are no constraints on $\left(Q/ M\right)$ for the E${\rm \overline{M}}$${\rm \overline{D}}$ black hole. In non of the two cases with phantom scalar field the photon sphere and the event horizon merge.

\subsection{Einstein Maxwell Dilaton black holes}

\begin{figure}[htbp]%
\begin{center}
\includegraphics[width=0.3\textwidth]{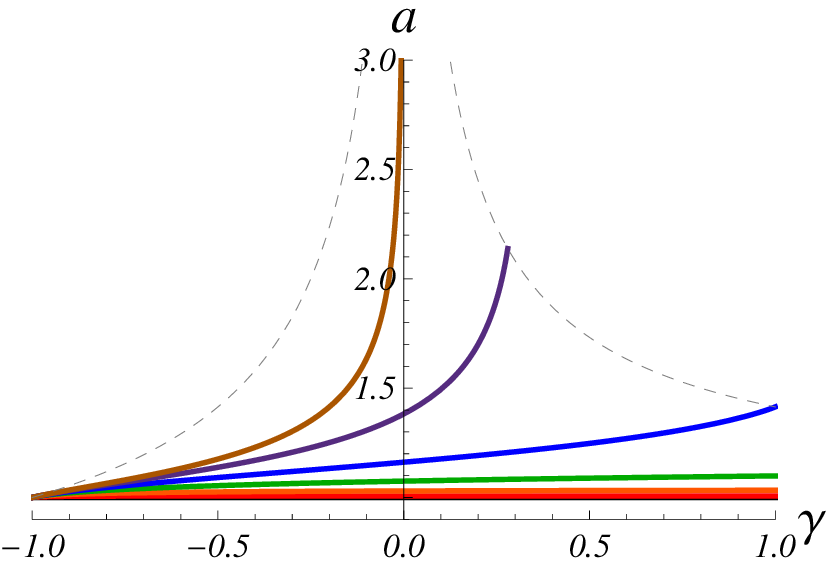}
\includegraphics[width=0.3\textwidth]{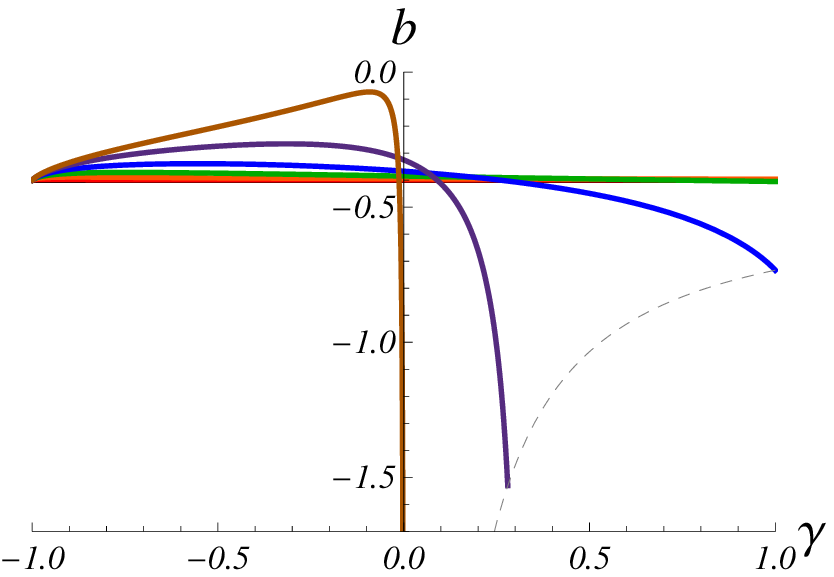}
\includegraphics[width=0.3\textwidth]{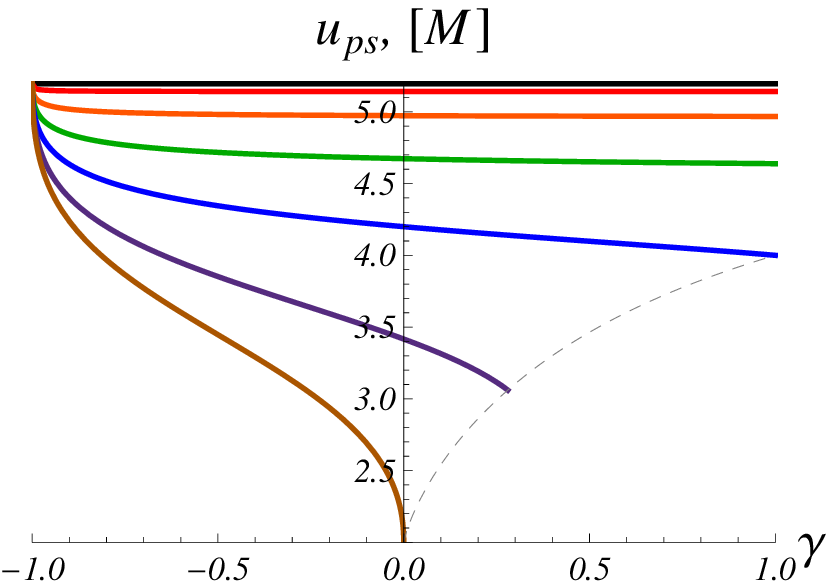}
\caption{\small
The EMD lens parameters $a$, $b$ and $u_{ps}$ for the following values of $Q/M$: $Q/M=0$ (black), $Q/M=0.25$ (red), $Q/M=0.5$ (orange),
$Q/M=0.75$ (green), $Q/M=1$ (blue), $Q/M=1.25$ (purple), $Q/M=\sqrt{2}$ (brown).} \label{lens_EMD}%
\end{center}
\end{figure}

\begin{figure}[htbp]%
\begin{center}
\includegraphics[width=0.3\textwidth]{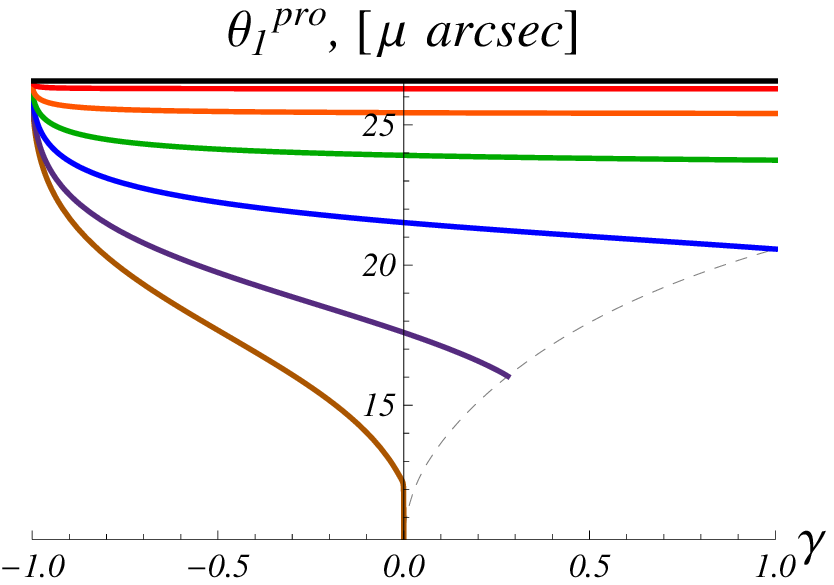}
\includegraphics[width=0.3\textwidth]{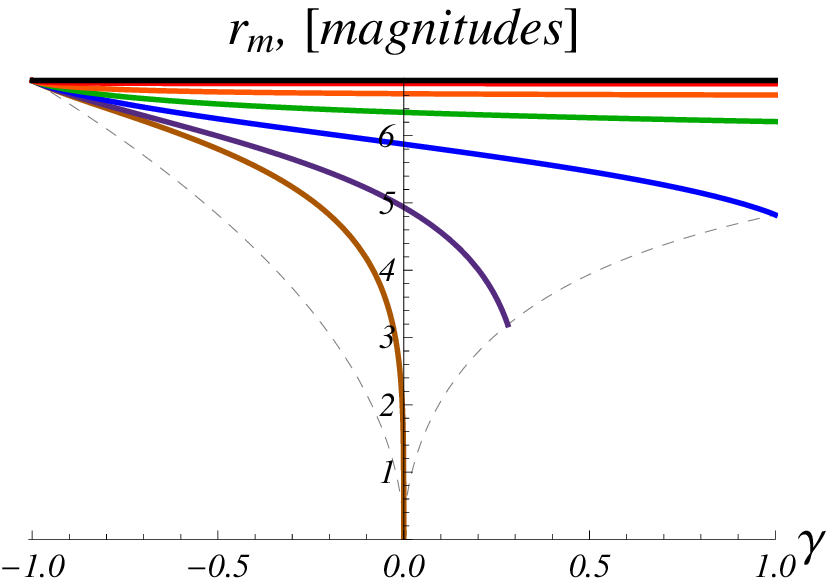}
\includegraphics[width=0.3\textwidth]{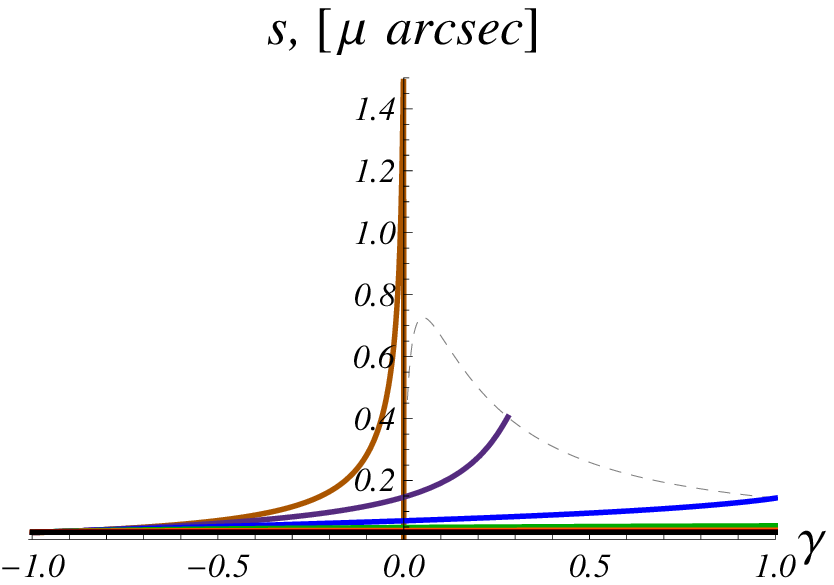}
\caption{\small
The observables $\theta_1^{pro}$, $r_m$ and $s_1^{pro}$ for the EMD black hole. The values of $Q/M$ are the same as on Fig. \ref{lens_EMD}. } \label{obs_EMD}%
\end{center}
\end{figure}

The lens parameters  $a$, $b$ and $u_{ps}$  for ${\rm EMD}$ case are given on Fig. \ref{lens_EMD}. The observables are given on Fig. \ref{obs_EMD}. The dashed line represents the critical curves of the parameters. For example, the critical curve for $a$ is defined as $a_{{\rm crit}}(\gamma)=a(\left(Q/ M\right)_{{\rm crit}},\gamma)$, where $\left(Q/ M\right)_{{\rm crit}}$ is the critical value of $\left(Q/ M\right)$ for the corresponding class of black-hole solutions. The critical curves of all other quantities in the paper are defined analogously and are represented by thin dashed lines. The regions beyond the critical curves on the figures correspond to naked singularities and are outside the scope of the current research.

In our discussion we will take the Schwarzschild black hole as a reference. The values of the different quantities corresponding to that case are presented by a straight grey line on the figures which we term ``reference line''.
The first observation we can make is that on both Fig. \ref{lens_EMD} and  Fig. \ref{obs_EMD} all curves converge to the value for the Schwarzschild
black hole at $\gamma=-1$ for arbitrary value of $Q/M$ -- a fact with no trivial explanation. The lens parameter $a$ is monotonous function of $\gamma$.
The slope is positive and becomes more significant with the increase of the electric charge $Q/M$. The parameter $b$ has a different behavior.
Initially it increases with $\gamma$ but then it passes through a maximum and then decreases. The branch with negative slope becomes very steep as
$Q/M$ is increased. Initially the EMD value of $b$ is higher than the Schwarzschild but for high enough values of $\gamma$ the situation changes.
The lowest value of $b$ is obtained at $\left(Q/ M\right)_{{\rm crit}}$ and $\gamma=0$ which is the value of the coupling in string theory \cite{GHS}.
For the EMD black holes the critical impact parameter $u_{ps}$ is lower than the Schwarzschild case for all non-zero values of $Q/M$. As $Q/M$ increase $u_{ps}$ decreases.
This effect, however, is compensated when stronger coupling and respectively lower value of $\gamma$ is considered.
%Due to the electromagnetic charge the images are closer to the black hole. Their brightness decreases with the increase of $Q/M$ and $\gamma$.
%The separation between them, however, increases as $Q/M$ and $\gamma$ are increased.

As we can see from Fig. \ref{obs_EMD} with the increase of $Q/M$ the relativistic images are attracted towards the black hole, they become less bright and the separation between them increases. The most demagnified image is obtained for $\left(Q/ M\right)_{{\rm crit}}$ and $\gamma=0$.
The dependence on $\gamma$ becomes more pronounced for higher values of $Q/M$. All three observables are monotonous functions of $\gamma$. The slope  of $\theta_1^{pro}$ and $r_m$ as functions of $\gamma$ are negative, while the slope of $s_1^{pro}$ is positive. In the case of stronger coupling the effect of the electric charge is suppressed. As a result, when $\gamma\rightarrow-1$ for all values of $Q/M$ the  relativistic images of the EMD black hole have the same angular position, brightness and separation as those of the Schwarzschild black hole.


\subsection{Einstein anti-Maxwell Dilaton black holes}

\begin{figure}[htbp]%
\begin{center}
\includegraphics[width=0.3\textwidth]{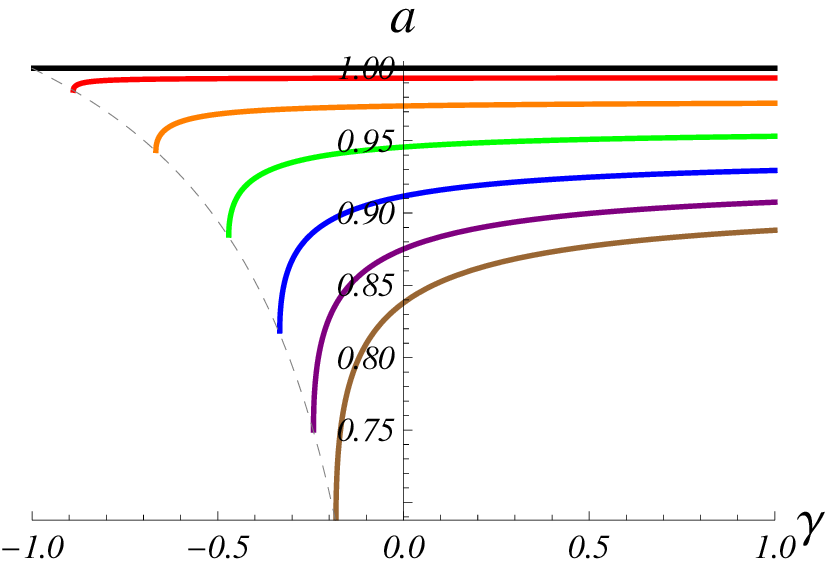}
\includegraphics[width=0.3\textwidth]{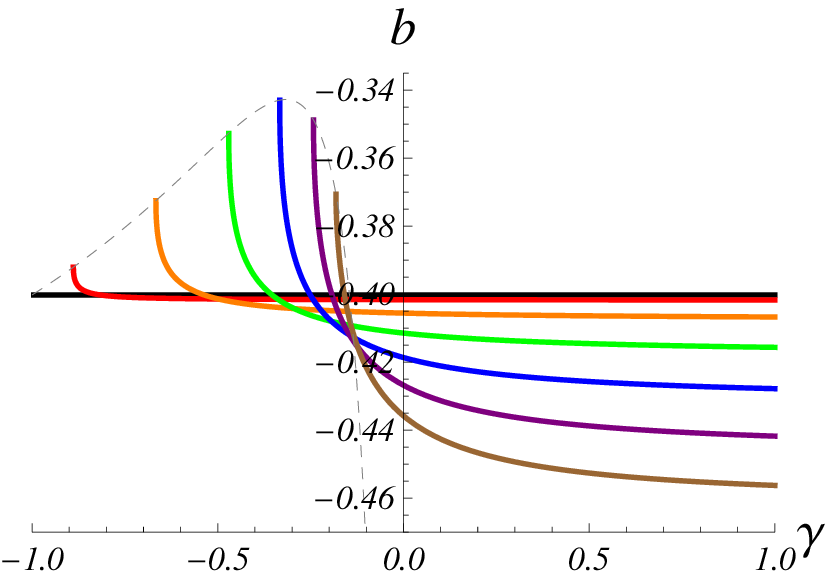}
\includegraphics[width=0.3\textwidth]{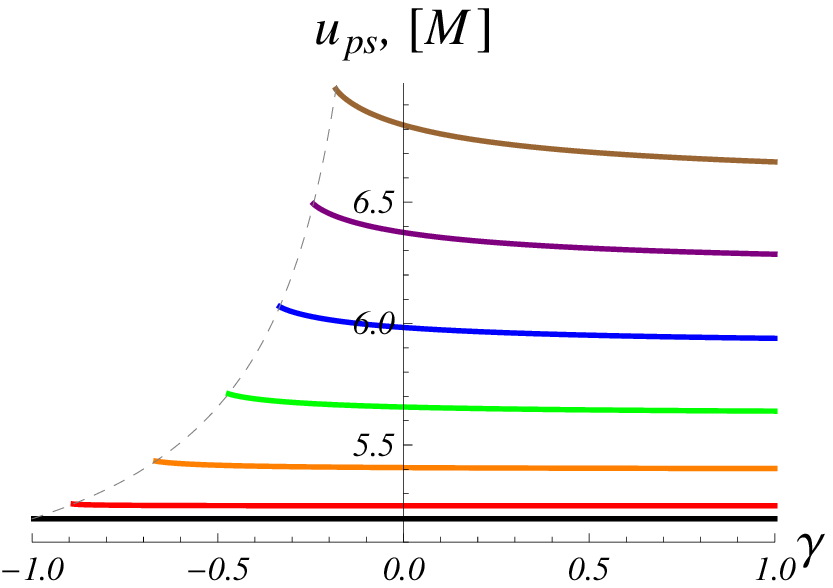}
\caption{\small The E$\overline{\rm M}$D lens parameters $a$, $b$ and $u_{ps}$ for the following values of $Q/M$: $Q/M=0$ (black), $Q/M=0.25$ (red), $Q/M=0.5$ (orange), $Q/M=0.75$ (green), $Q/M=1$ (blue), $Q/M=1.25$ (purple), $Q/M=1.5$ (brown).} \label{lens_E_MD}%
\vspace{8mm}
\includegraphics[width=0.3\textwidth]{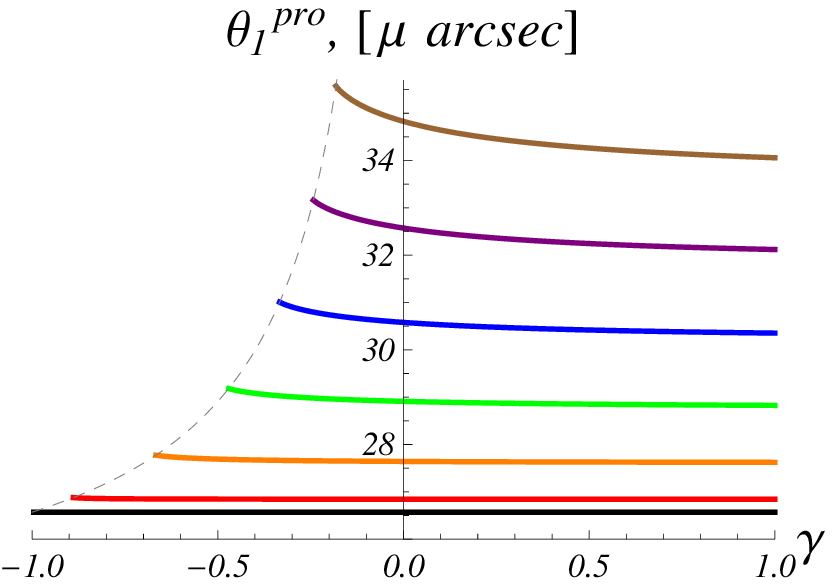}
\includegraphics[width=0.3\textwidth]{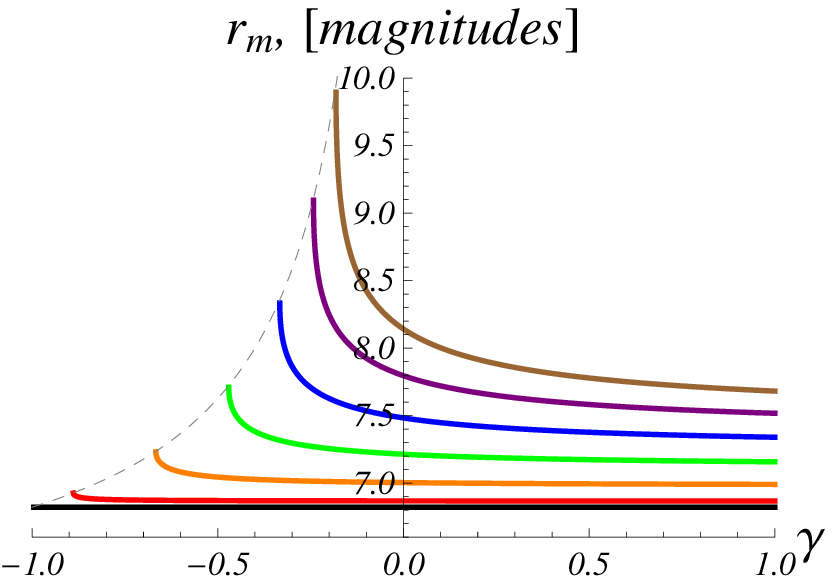}
\includegraphics[width=0.3\textwidth]{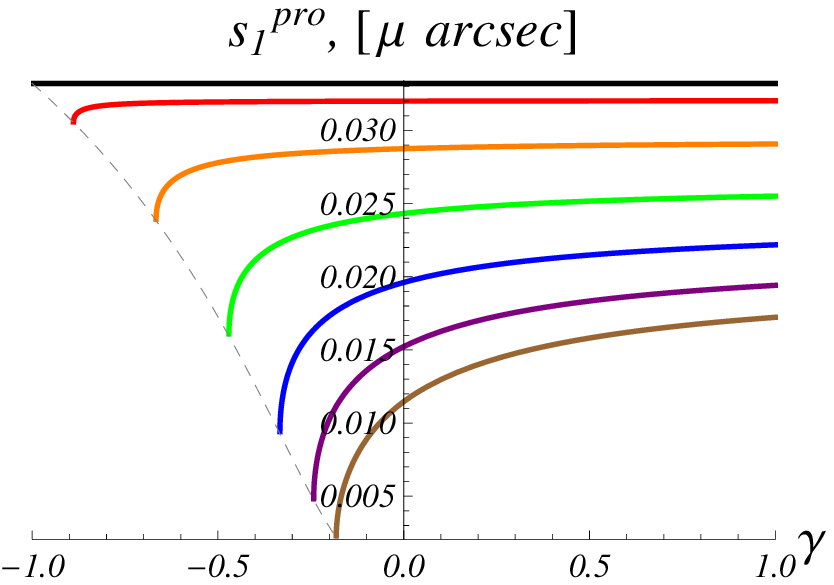}
\caption{\small The observables $\theta_1^{pro}$, $r_m$ and $s_1^{pro}$ for the E$\overline{\rm M}$D black hole.
The values of $Q/M$ are the same as on Fig. \ref{lens_E_MD}. } \label{obs_E_MD}%
\end{center}
\end{figure}

The results for the ${\rm E\overline{M}D}$ case are presented on Fig. \ref{lens_E_MD} and  Fig. \ref{obs_E_MD}.
On all of the graphics for the ${\rm E\overline{M}D}$ black hole the curves end on the critical curves, before the value $\gamma=-1$ is reached. Beyond the critical curves the object is not a black hole anymore.

Unlike the previously discussed case, the lens parameter $a$ decreases when $Q/M$ is increased. $a$ is monotonous function of $\gamma$ and as in the EMD case the slope of the curves is positive. Here the stronger coupling enhances the effect of the electric charge. In the ${\rm E\overline{M}D}$ case $b$ is monotonous function of $\gamma$ but its behavior is again more complex than that of $a$. The slope of the curves is negative. For high enough values of $\gamma$ with the increase of $Q/M$, $b$ decreases. With the decrease of $\gamma$, however, the curves cross the reference line and the value of $b$ becomes higher than that for the Schwarzschild black hole. The behavior of $u_{ps}$ is converse to that of $a$ -- higher charge leads to higher values. The effect of the charge is enhanced when the coupling is stronger.

What are the effect of the phantom electromagnetic field and the dilaton on the observables? The effect of the phantom electric charge is to repel the relativistic images from the optical axis. The slope of the curves for $\theta_1^{pro}$ is negative. It is negligible for low values of $Q/M$. The stronger coupling leads to a more pronounced effect of the phantom electric charge. The qualitative behavior of the curves for $r_m$ is identical but the curves are much steeper. The separation between the images $s_1^{pro}$ has a converse behavior. It is lower for the images that
a farther from the optical axis. The slope of $s_1^{pro}$ is positive.
%In this case the effect of the
%electromagnetic charge is to repel the images from the black hole. They become brighter but the separation between them is decreased. These effects
%become more appreciable as $\gamma$ is decreased.
\vspace{8mm}

\subsection{Einstein Maxwell anti-Dilaton black holes}

\begin{figure}[htbp]%
\begin{center}
\includegraphics[width=0.3\textwidth]{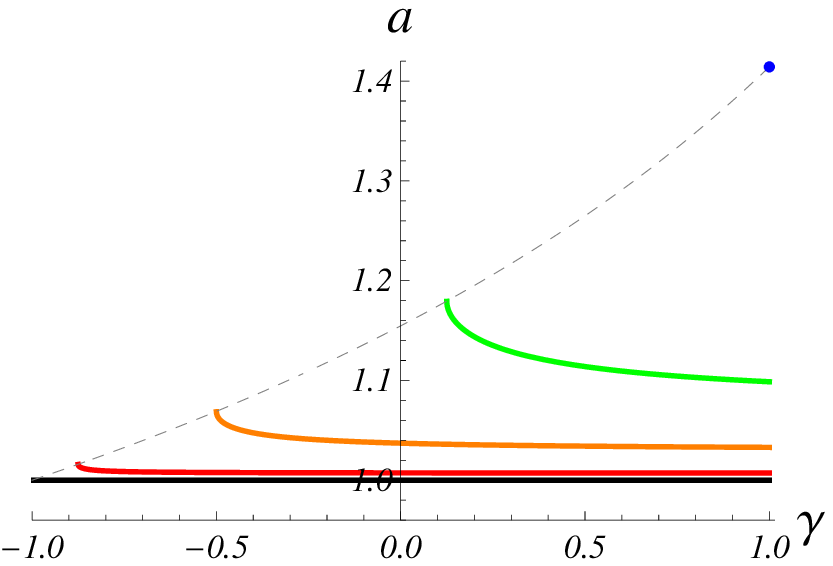}
\includegraphics[width=0.3\textwidth]{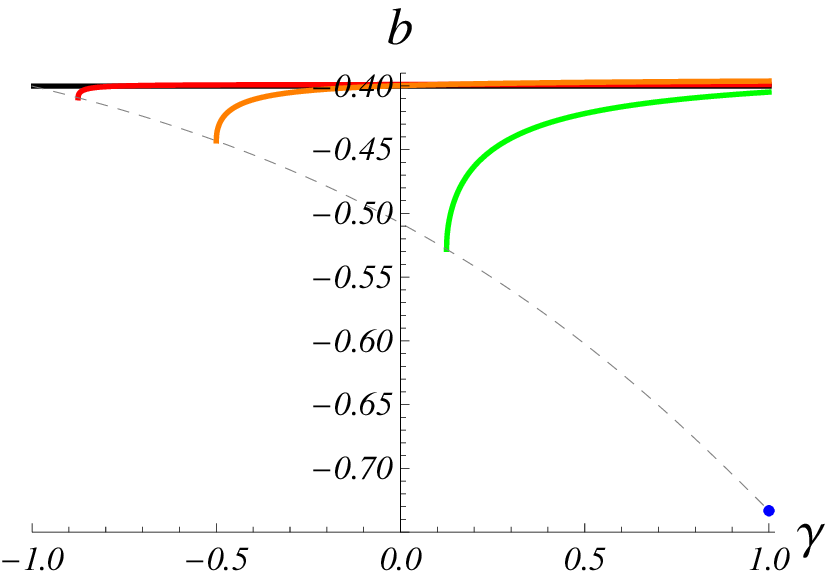}
\includegraphics[width=0.3\textwidth]{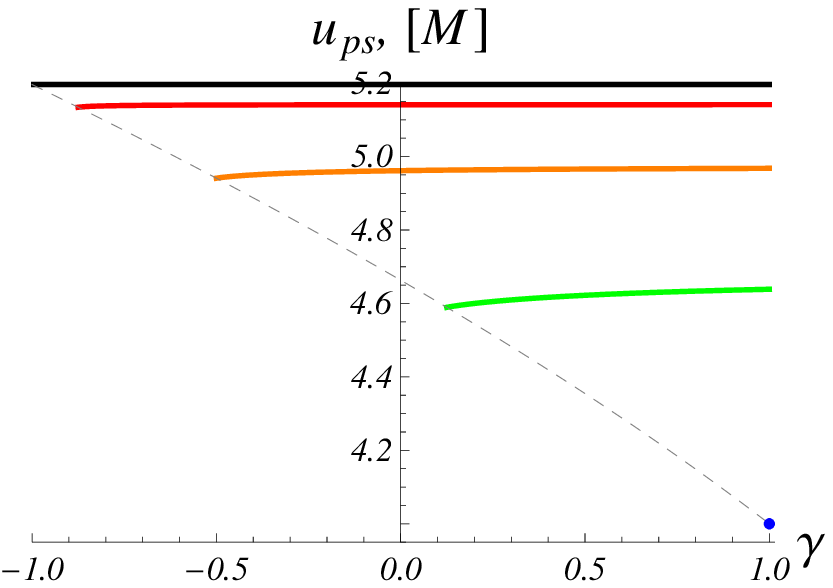}
\caption{%
\small The EM$\overline{\rm D}$ lens parameters $a$, $b$ and $u_{ps}$ for the following values of $Q/M$: $Q/M=0$ (black), $Q/M=0.25$ (red), $Q/M=0.5$ (orange),
$Q/M=0.75$ (green), $Q/M=1$ (blue).} \label{lens_EM_D}%
\vspace{11mm}
\includegraphics[width=0.3\textwidth]{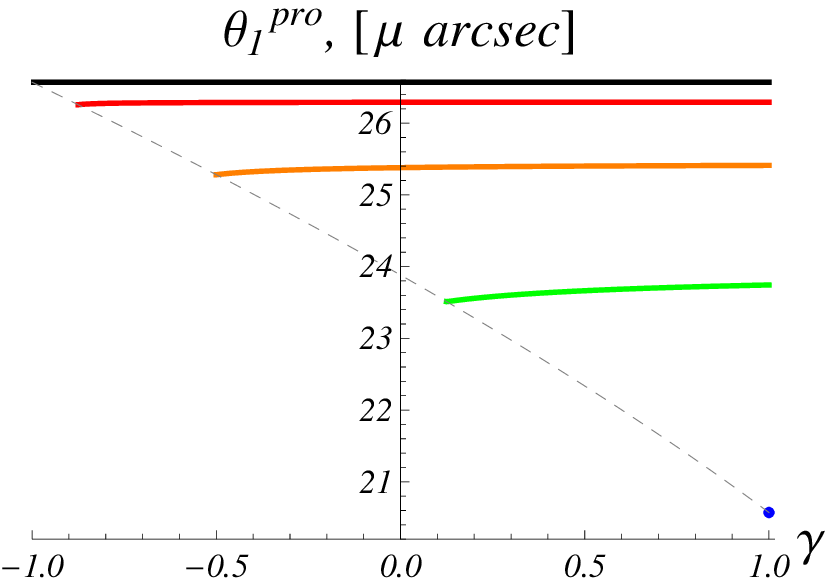}
\includegraphics[width=0.3\textwidth]{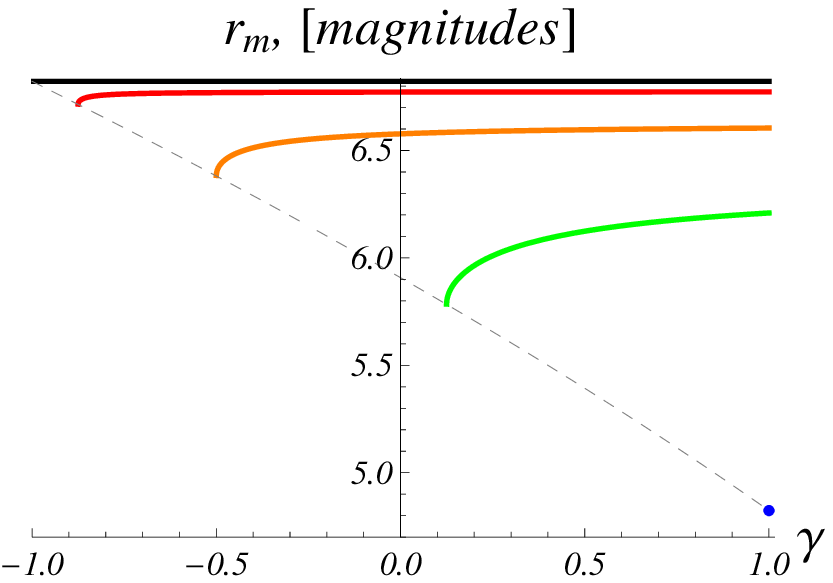}
\includegraphics[width=0.3\textwidth]{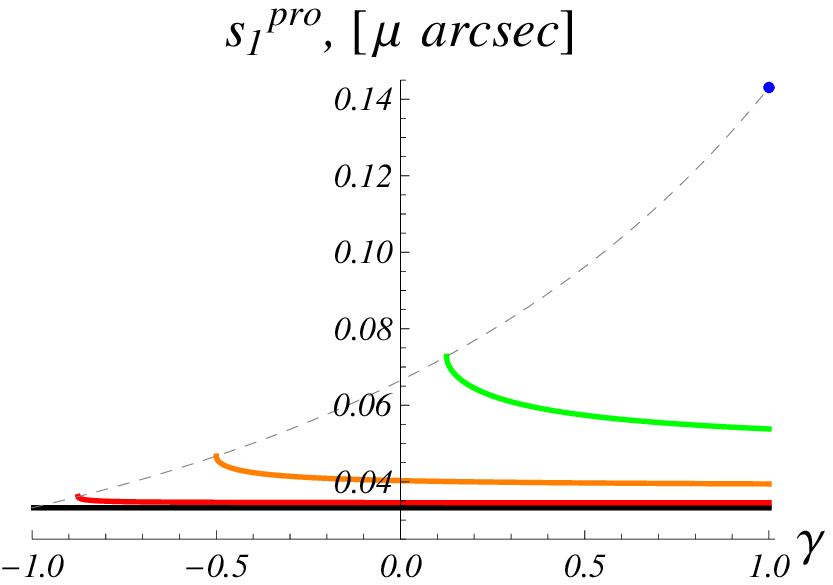}
\caption{%
\small The observables $\theta_1^{pro}$, $r_m$ and $s_1^{pro}$ for the EM$\overline{\rm D}$ black hole.
The values of $Q/M$ are the same as on Fig. \ref{lens_EM_D}.} \label{obs_EM_D}%
\end{center}
\end{figure}

The lens parameters and the observable in the case of  EM$\overline{\rm D}$ are presented on Fig. \ref{lens_EM_D} and  Fig. \ref{obs_EM_D}, respectively.
Here again the point $\gamma=-1$ is reached only when $Q/M=0$. Otherwise the curves end on the critical lines.
The lens parameter $a$ is a monotonous function of $\gamma$. For $Q/M\neq0$ it is higher than the reference value.
The negative slope here means that the effect of the electromagnetic field is enhanced when $\gamma$ is decreased.
The behavior of $b$ and $u_{ps}$ is converse -- they decrease as $Q/M$ is increased.
For $b$ the effect of a stronger coupling is to invoke a stronger effect of $Q/M$.
The critical impact parameter is almost independent of $\gamma$ as we can see from the almost flat curves. As a result of that,
the value of the angular position of the images $\theta_1^{pro}$ is also slightly dependent on $\gamma$.
The images are attracted to the optical axis with the increase of $Q/M$. They become less bright as the electric charge is increased and
this effect is more significant for higher coupling. The behavior of $s$ is converse -- it is higher for higher $Q/M$
%With the increase of electric charge $Q/M$ the angular position  and the magnification of the images decrease while the separation
%between them increases. The angular positions seem to be slightly dependent on $\gamma$ while the behavior of the two later observables
%becomes more appreciable as $\gamma$ appoaches $-1$.
\vspace{5mm}

\subsection{Einstein anti-Maxwell anti-Dilaton black holes}

\begin{figure}[t]%
\begin{center}
\includegraphics[width=0.3\textwidth]{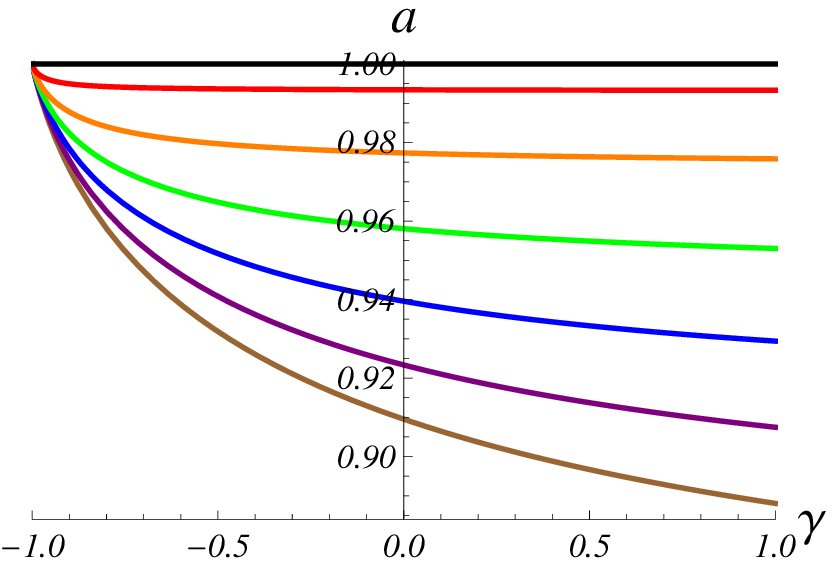}
\includegraphics[width=0.3\textwidth]{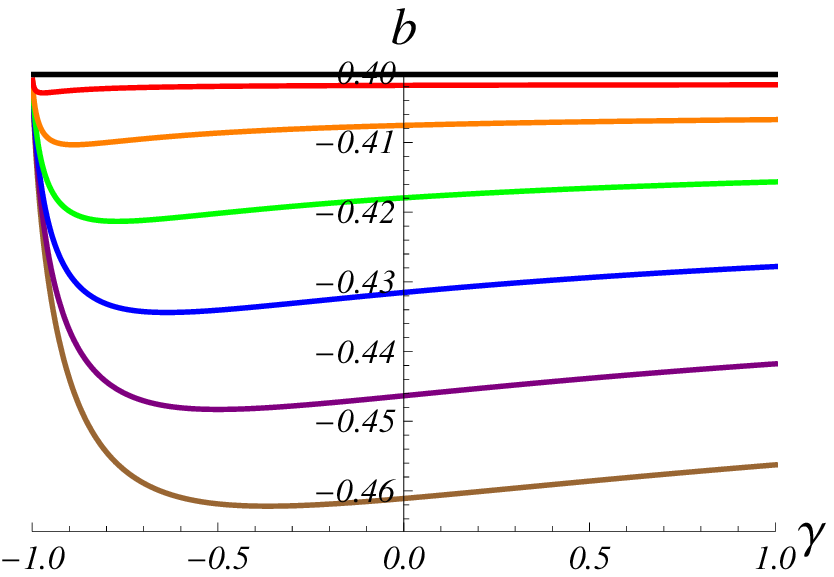}
\includegraphics[width=0.3\textwidth]{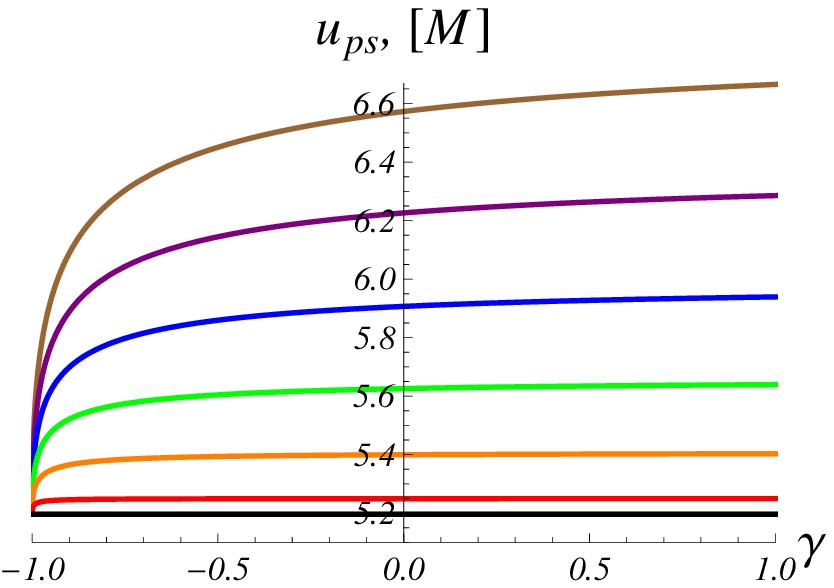}
\caption{%
\small The E$\overline{\rm M}$$\overline{\rm D}$ lens parameters $a$, $b$ and $u_{ps}$ for the following values of $Q/M$: $Q/M=0$ (black), $Q/M=0.25$ (red), $Q/M=0.5$ (orange),
$Q/M=0.75$ (green), $Q/M=1$ (blue), $Q/M=1.25$ (purple), $Q/M=1.5$ (brown).} \label{lens_E_M_D}%
\vspace{5mm}
\includegraphics[width=0.3\textwidth]{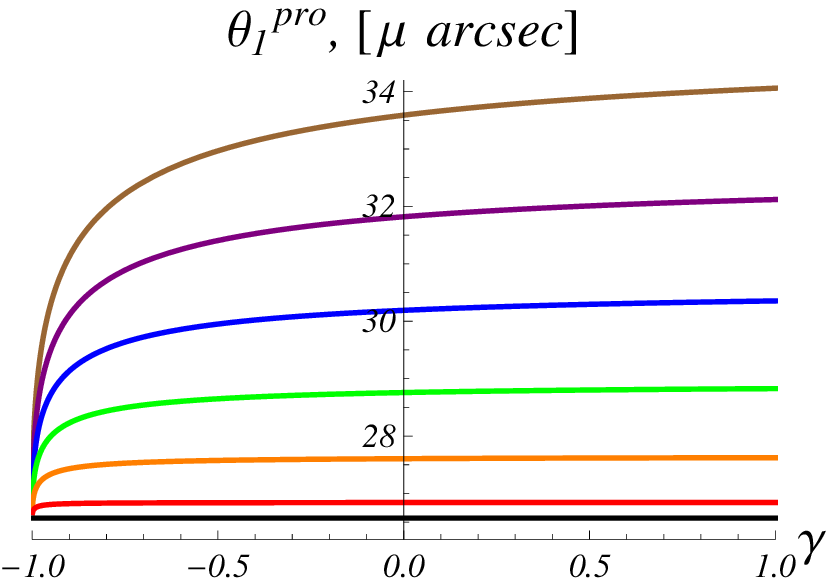}
\includegraphics[width=0.3\textwidth]{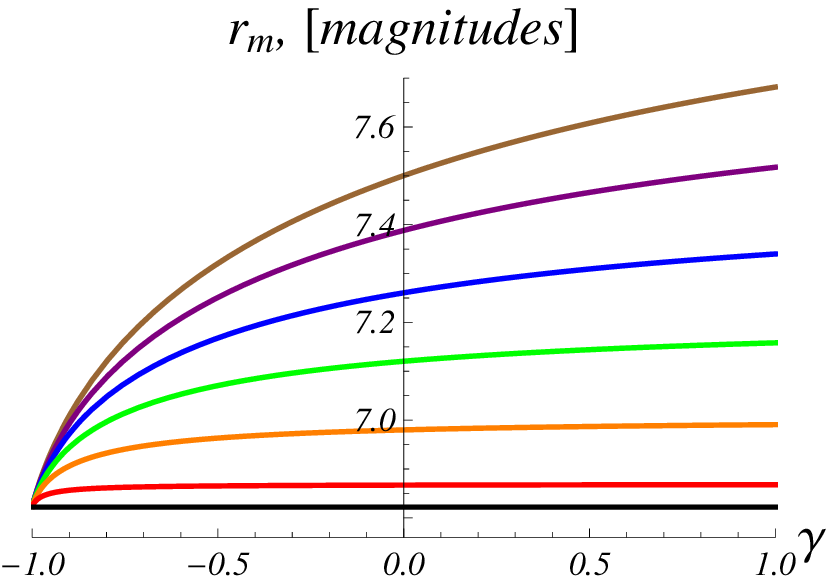}
\includegraphics[width=0.3\textwidth]{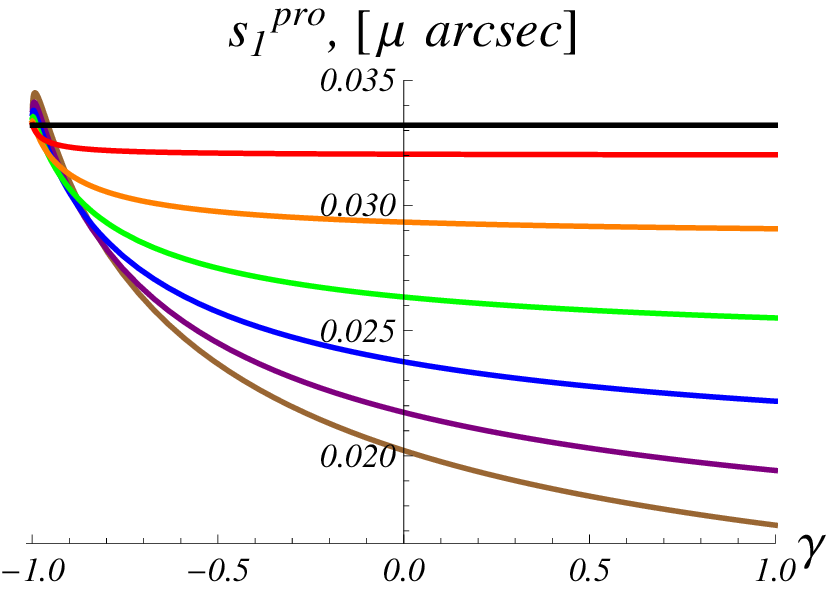}
\caption{%
\small The observables $\theta_1^{pro}$, $r_m$ and $s_1^{pro}$ for the E$\overline{\rm M}$$\overline{\rm D}$  black hole.
The values of $Q/M$ are the same as on Fig. \ref{lens_E_M_D}.} \label{obs_E_M_D}%
\end{center}
\end{figure}

Fig. \ref{lens_E_M_D} and  Fig. \ref{obs_E_M_D} represent the results for the last case -- the E$\overline{\rm M}$$\overline{\rm D}$ black hole. As we mentioned above, in this case there are no restrictions for the electric charge so no critical curves occur on the graphics. Here, just as in the EMD case, all curve converge to the Schwarzschild line when $\gamma=-1$. The lens parameter $a$ is lower when $Q/M$ is increased. The effect of the phantom electric field, however, is suppressed in the strong coupling regime. Again, $b$ is not monotonous. It is lower than the Schwarzschild value for all values of $Q/M\neq0$ and $\gamma\neq-1$. With the decrease of $\gamma$, $b$ initially decreases. Then, it passes through a minimum and converges to the reference line. The negative slope becomes more steep with the increase of $Q/M$. The critical impact parameter $u_{ps}$ has behavior opposite to that of $a$. It is higher for higher values of $Q/M$. Its dependence on $\gamma$ is insignificant for high enough values of $\gamma$ but the curves become very steep as the point $\gamma=-1$ is approached.

Due to the phantom electromagnetic field the relativistic image are observed at higher angular position $\theta_1^{pro}$. The dependence of $\theta_1^{pro}$ on $\gamma$ is almost negligible everywhere but in the vicinity of $\gamma=-1$. Again the images that are observed farther from the optical axis are also brighter. The slope of the curve for $r_m$ is bigger than that of the previous graphic when equal values of $Q/M$ are considered. The separation between the first and second relativistic images $s_1^{pro}$  has odd behavior. For low values of $Q/M$ it is a monotonous function of $\gamma$. For decreasing $\gamma$, $s$ increases. For high enough values of $Q/M$ as $\gamma$ is decreased the curves for $s$ cross the reference line and becomes higher than the value for Schwarzschild. Then it has a local maximum and finishes on the Schwarzschild line at $\gamma=-1$.
%With the increase of electric charge $Q/M$ the angular position  and the magnification of the images increase while the separation
%between them decreases. Unlike the two previous cases the effect of the electric charge $Q/M$ becomes less appreciable as $\gamma$ appoaches $-1$.

\section{Comparison between the four cases and summary of the results}

\begin{figure}[htbp]%
\begin{center}
\includegraphics[width=0.4\textwidth]{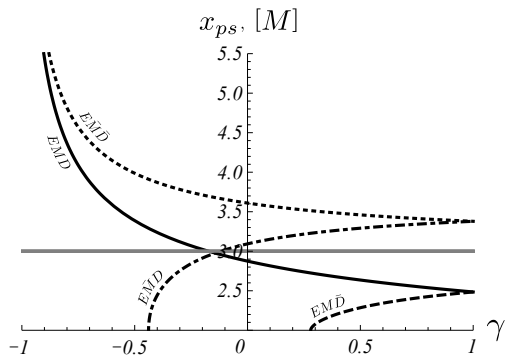}
\caption{%
\small The photon sphere for $Q/M=0$ corresponding to the Schwarzschild black hole (grey) and $Q/M=0.8$  for the other four cases --
EMD (thick), E$\overline{\rm M}$D (dash-dot), EM$\overline{\rm D}$ (dash), E$\overline{\rm M}$$\overline{\rm D}$ (dot).} \label{PhotonSpheres_all}%
\end{center}
\end{figure}

\begin{figure}[t]%
\begin{center}
\includegraphics[width=0.4\textwidth]{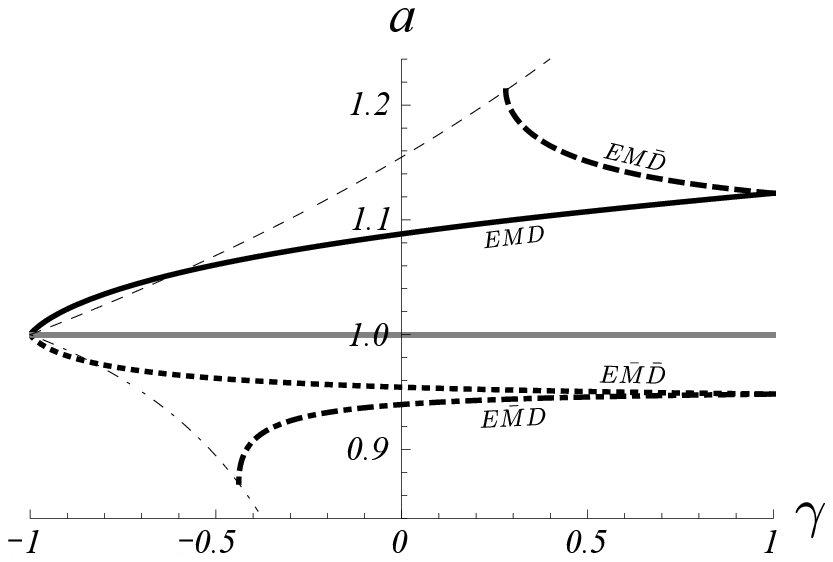}
\includegraphics[width=0.4\textwidth]{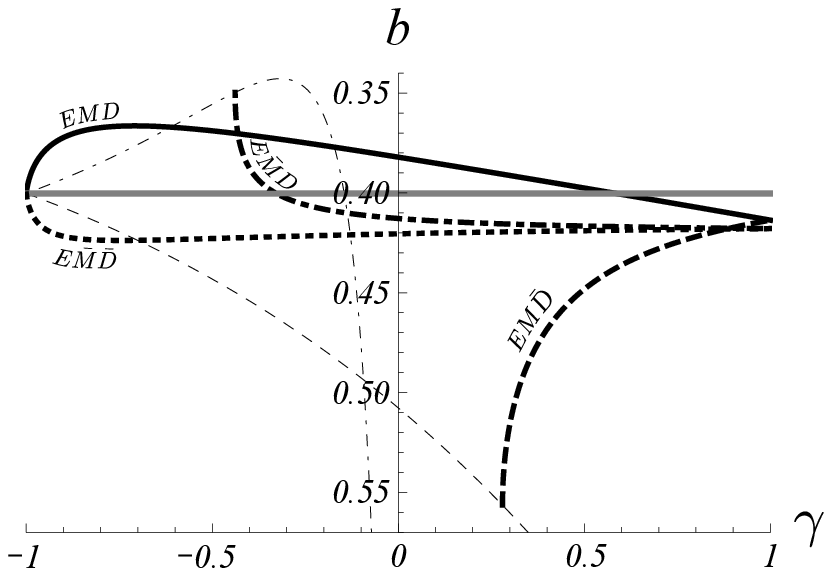}
\caption{%
\small The lens parameters $a$ and $b$ for $Q/M=0$ corresponding to the Schwarzschild black hole (grey) and $Q/M=0.8$  for the other four cases --
EMD (thick), E$\overline{\rm M}$D (dash-dot), EM$\overline{\rm D}$ (dash), E$\overline{\rm M}$$\overline{\rm D}$ (dot).} \label{a_b_all}%
\end{center}
\end{figure}

\begin{figure}[htbp]%
\begin{center}
\includegraphics[width=0.4\textwidth]{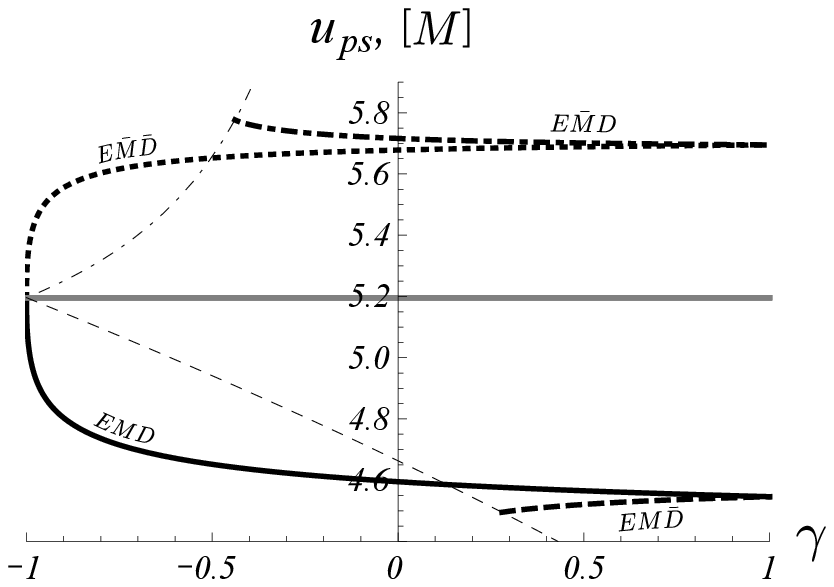}
\includegraphics[width=0.4\textwidth]{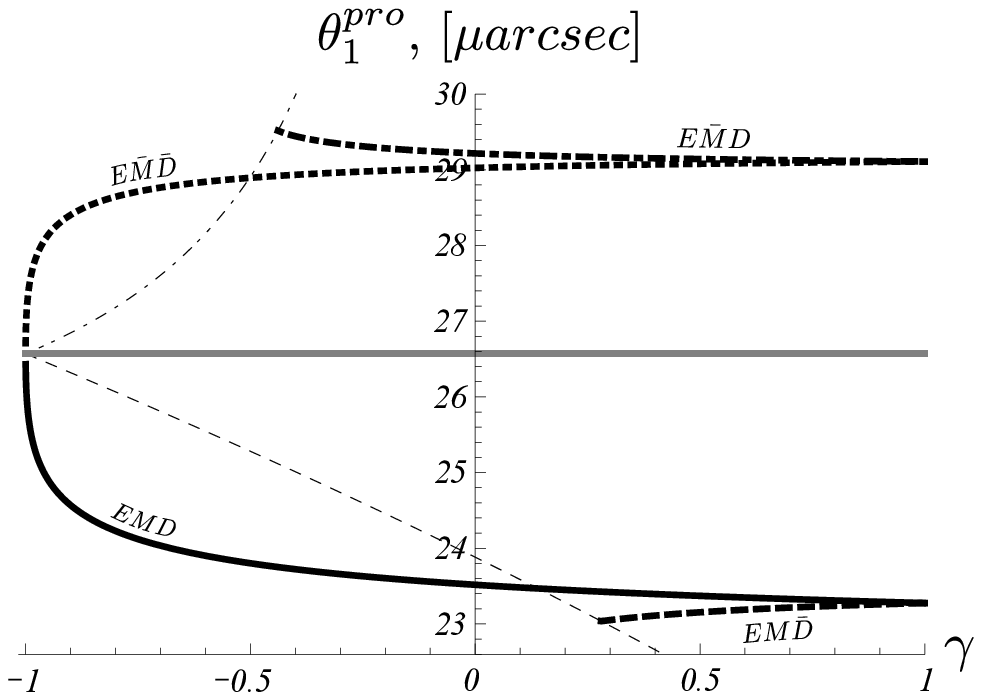}
\caption{%
\small The impact parameter $u_{ps}$ and the angular position of the first relativistic image for prograde photons $\theta_{1}^{pro}$
for $Q/M=0$ corresponding to the Schwarzschild black hole (grey) and $Q/M=0.8$  for the other four cases --
EMD (thick), E$\overline{\rm M}$D (dash-dot), EM$\overline{\rm D}$ (dash), E$\overline{\rm M}$$\overline{\rm D}$ (dot).} \label{ups_theta_all}%
\end{center}
\end{figure}

\begin{figure}[htbp]%
\begin{center}
\includegraphics[width=0.4\textwidth]{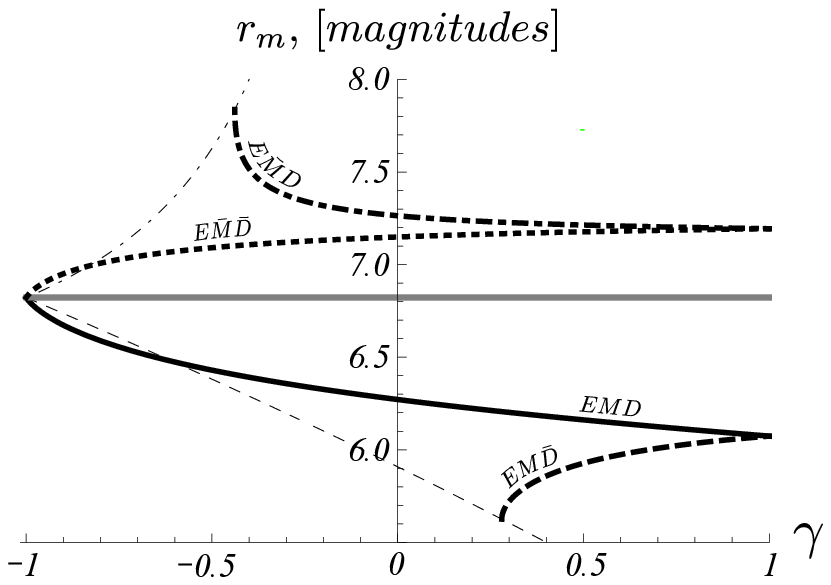}
\includegraphics[width=0.4\textwidth]{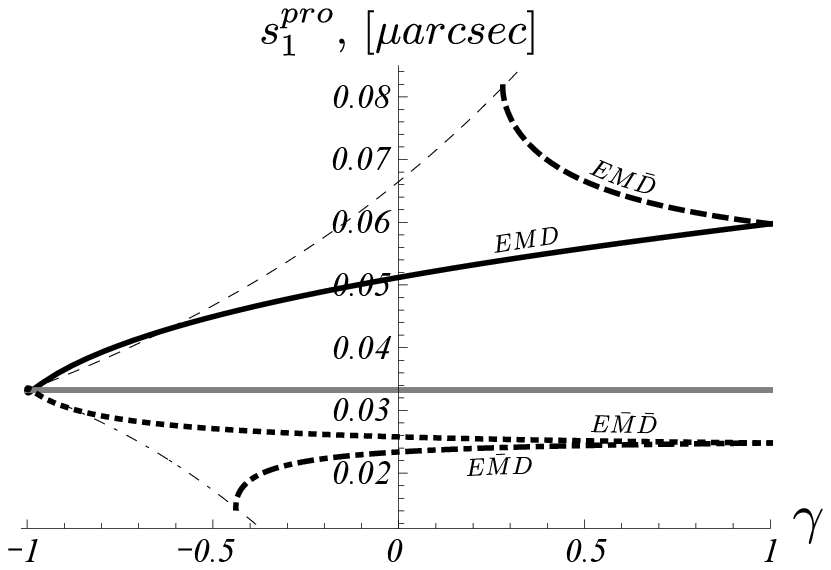}
\caption{%
\small The flux ratio $r_m$ and the angular separation between the first and second relativistic image for prograde photons $s_1^{pro}$
for $Q/M=0$ corresponding to the Schwarzschild black hole (grey) and $Q/M=0.8$  for the other four cases --
EMD (thick), E$\overline{\rm M}$D (dash-dot), EM$\overline{\rm D}$ (dash), E$\overline{\rm M}$$\overline{\rm D}$ (dot).} \label{rm_s_pro_1 _all}%
\end{center}
\end{figure}

In this section we will compare between the four cases -- (${\rm EMD}$), (${\rm E\overline{M}D}$), (${\rm EM\overline{D}}$) and (${\rm E\overline{ M}\overline{ D}}$) -- for black holes with same mass $M$ and electric charge $Q$. For all of the discussed cases on the same plot the photon sphere  $x_{ps}$  is presented on Fig. \ref{PhotonSpheres_all}, the lens parameters  $a$ and $b$ --  on Fig. \ref{a_b_all}, the impact parameter $u_{ps}$ and the angular position $\theta_1^{pro}$ are on Fig. \ref{ups_theta_all}, and the other two observables, $r_m$ and $s_1^{pro}$, are given on Fig. \ref{rm_s_pro_1 _all}. On all graphics in the current section $M=1$ and $Q=0.8$. For most of the quantities the curves corresponding to black hole with canonical electromagnetic field lay on one side of the reference line while those corresponding to phantom electromagnetic field -- on the other. Exception from this behavior is observed for the photon sphere $x_{ps}$ and for the lens parameter $b$.

%The photon sphere plays a significant role for the formation of relativistic images. In all four cases it is present for the entire range of admissible values of the parameters -- $M$, $Q$ and $\gamma$.

For weak coupling ($\gamma$ close to $1$) both black holes with canonical electromagnetic field ${\rm EMD}$ and ${\rm EM\overline{D}}$ have
photon spheres with smaller radii than the Schwarzschild black hole while the black holes with phantom electromagnetic field have bigger radii.
The situation changes when $\gamma$ is decreased. The curve for ${\rm EMD}$ case does not remain below the reference line but crosses it and
diverges as $\gamma=-1$ is approached. The curve for ${\rm E\overline{M}D}$ case also crosses the reference line but downwards and disappears
when the critical value of $\gamma$ corresponding to $Q/M=0.8$ is reached. It is important to note that in none of the cases the photon sphere converges to the reference line in the limit $\gamma=-1$.

%Let us first compare between the effects of the Maxwell field and anti-Maxwell field on the relativistic images. The Maxwell field attracts the images. Due to the electromagnetic field they are closer to the black hole in comparison to the Schwarzschild case and less bright. The separation between the images of different order, the first and the second relativistic image in particular, however, is increased. The phantom (anti-Maxwell) electromagnetic field has an opposite effect. The images are repelled from the black hole and brighter but the separation between them is decreased.
%In both cases the effects become more significant with the increase of the electromagnetic charge and remain valid for all values of the parameter $\gamma$.

As we can see from Fig. \ref{a_b_all} for both black holes with canonical electromagnetic field, ${\rm EMD}$ and ${\rm EM\overline{D}}$, $a$ is higher than the Schwarzschild value in the whole interval of admissible values of $\gamma$, while
for the solutions with phantom electromagnetic field, ${\rm E\overline{M}D}$ and ${\rm E\overline{ M}\overline{ D}}$, it is lower.

What is the role of the parameter $\gamma$ responsible for the coupling between the dilaton and the Maxwell field? Let us first consider the
couple of black hole solutions with canonical scalar field ${\rm EMD}$ and ${\rm E\overline{M}D}$. As it can be seen from Fig. \ref{a_b_all} for lower values of $\gamma$, corresponding
to stronger coupling, $a$ has lower values. In the  phantom scalar field case (see the curves for the ${\rm EM\overline{D}}$ and the ${\rm E\overline{ M}\overline{ D}}$ solutions)
on the contrary -- the stronger coupling leads to higher values of $a$.

As a result, for the ${\rm EMD}$ and ${\rm E\overline{ M}\overline{ D}}$ black holes the stronger coupling suppresses the effect of the Maxwell field and the curves for $a$ converge to the reference line corresponding to the Schwarzschild black hole,
while for the ${\rm E\overline{M}D}$ and ${\rm EM\overline{D}}$ black holes the effects of the two parameters $Q/M$ and $\gamma$ enhance each other and the curves
diverge from the reference line.

The curves for the lens parameter $b$ have a more complex behavior. At $\gamma=1$ for $Q/M\neq0$ for all four of the considered black holes
$b$ has lower values than for the Schwarzschild black hole. As the coupling is increased (and respectively $\gamma$ is decreased) for both
solutions with canonical scalar field the curves cross the reference line and $b$ takes higher values. For the case of phantom scalar field in the
whole interval of admissible values of $\gamma$ the values of $b$ remain lower than those of the Schwarzschild case. As for the previously
discussed parameter the curves for $b$ in the  ${\rm EMD}$ and ${\rm E\overline{ M}\overline{ D}}$ cases converge to the Schwarzschild line at
$\gamma=-1$. In these cases $b$ has one extremum -- a maximum in the former case and a minimum in the latter case.

The qualitative behavior of the curves for the impact parameter $u_{ps}$ and for the  observables $\theta_{1}^{pro}$, $r_m$ and $s_1^{pro}$ is similar to that of the curves for $a$ in a sense that for the ${\rm EMD}$ and ${\rm E\overline{ M}\overline{ D}}$ black holes the curves  converge to corresponding Schwarzschild values, while for the other couple of black holes, ${\rm E\overline{M}D}$ and ${\rm EM\overline{D}}$, the curves diverge from them. All of these quantities are monotonous functions of $\gamma$.


For all four solutions the following behavior is observed. Images that are closer to the optical axis are dimmer but better separated while those that are farther -- on the contrary. In all cases $s_1^{pro}$ has a converse behavior to that of $\theta_{1}^{pro}$ and $r_m$ --  when the latter increase the former decreases.

From the studied cases we can conclude that the canonical electromagnetic field attracts the relativistic images towards the optical axis while the phantom electromagnetic field repels them. In the case of canonical scalar field the higher coupling repels the images form the optical axis while in the phantom scalar field case it attracts them.

In the limit of infinitely strong coupling for any value of $Q/M$ the ${\rm EMD}$ and the ${\rm E\overline{ M}\overline{ D}}$ black holes become practically indistinguishable from the Schwarzschild black hole on the bases of observations for the angular position, the magnification and the separation of the relativistic images.

%Now let us compare between the canonical and phantom scalar fields. From the numerical results presented above we can conclude that due to the presence of the phantom scalar field the images are attracted to the horizon, demagnified and the separation between them is increased. The canonical scalar field yields an opposite effect.

%As we can see in the different cases the images are either attracted to or repelled from the black hole.
%The numerical results do not show any tendency of this type for the photon sphere, however.

\vspace{1.5ex}
\begin{flushleft}
\large\bf Acknowledgments
\end{flushleft}

Partial financial support from the Bulgarian National Science Fund under
Grant DMU 03/6 is gratefully acknowledged. The authors would like to thank prof. S. Yazadjiev for the fruitful discussions and the anonymous referee for the valuable remarks.

\end{document}